\documentstyle[12pt,a4,epsfig]{article}
\newcommand{\be}{\begin{equation}}
\newcommand{\ee}{\end{equation}}
\newcommand{\ba}{\begin{eqnarray}}
\newcommand{\ea}{\end{eqnarray}}
\newcommand{\baa}{\begin{eqnarray*}}
\newcommand{\eaa}{\end{eqnarray*}}
\newcommand{\bb}{\begin{thebibliography}}
\newcommand{\eb}{\end{thebibliography}}

\newcommand{\bi}[1]{\bibitem{#1}}
\begin {document}

\rightline{ TPR-99-08}

\begin{center}
{\Large \bf Tensor polarization of vector mesons from quark and gluon
fragmentation}\\
\vspace*{1cm}
 A. Sch\"afer$^1$, L. Szymanowski$^{1,2}$, O.V. Teryaev$^{1,3}$ \\
\vspace*{0.3cm}
{$^1$\it Institut f\"ur Theoretische Physik\\
Universit\"at Regensburg\\
D-93040 Regensburg, Germany }\\
\vspace*{0.3cm}
{$^2$\it Soltan Institute for Nuclear Studies\\
Hoza 69, 00-681 Warsaw, Poland}\\
\vspace*{0.3cm}
{$^3$\it Bogoliubov Laboratory of Theoretical Physics\\
Joint Institute for Nuclear Research, Dubna\\
141980 Russia}\\
\vspace*{0.3cm}
\end{center}
\begin{abstract}

We considered the electro- and photoproduction
of $\rho$ and other vector mesons with moderately large 
transverse momentum in the scattering
of unpolarized electrons on unpolarized nucleons and nuclei. 
For these processes we have analyzed how to extract 
the novel fragmentation functions $\bar b_1^q(z), \bar b_1^G(z)$,
describing tensor polarization, 
and the photon double spin-flip distribution  
$\Delta^\gamma$
from the angular distribution
of meson decay products.

\end{abstract}

\newpage

Deep Inelastic Scattering (DIS) on targets with higher spin \cite{HJM89}
offers the possibility to study new spin structure
functions. The simplest one is the tensor spin structure function
$b_1$, appearing for target hadrons with spin-1 or higher
\cite{HJM89,FS}, and being the subject of recent model studies \cite{b1} and 
lattice calculations \cite{lat}. 

%Another interesting example is the structure function $\Delta$
%related to the hadronic double spin flip amplitude \cite{JM89} 
%(the relevant operators were identified earlier
%in \cite{BKLF85}).  

Another interesting example is the structure function $\Delta$
related to the amplitude which describes a 
spin flip of both the hadron and the photon by two units each \cite{JM89} 
(the relevant operators were identified earlier
in \cite{BKLF85}).  

For a photon target, it is proportional to the structure function
$F_3^\gamma$, whose small-$x$ behaviour was studied recently in
\cite{EKS99}.

For practical experiments the only available hadronic target with spin
1 is the deuteron. 
Since both   $\Delta$ and $b_1$ are zero, when neutron and proton are
scattered independently, the observable effects are expected to be 
very small. The reason is the weakness of the nuclear binding, as
the tensor polarization of any spin-one particle built from 
two independent spin-$1/2$ constituents is zero. 

At the same time, final states including vector mesons are readily
produced in hard hadronic and leptonic processes. 
As quarks in the
vector mesons are strongly bound, 
the latter  can have a large tensor polarization (it is a genuine
spin-$1$
object in contrast to the deuteron), 
which indeed was
observed experimentally, see e.g. \cite{exp}. This observation suggests, that 
the proper places to observe effects of $b_1$ and $\Delta$ are 
their fragmentation counterparts. This is the main subject 
of the present paper. 

We identify the fragmentation analogs of $b_1$ and $\Delta$ which
parametrize the tensor polarization of vector
mesons, which can be experimentally accessed by the observation of
specific angular distributions of their decay products. 

By simultaneously observing high-$P_T$ vector meson production 
at large and small photon virtualities $Q^2$ at HERA one could 
measure new fragmentation functions as well as the (resolved) photon 
distribution
$\Delta^\gamma$, directly related to the 
structure function $F_3^\gamma$.
   
We consider the vector meson produced in fragmentation processes of
either 
a leading
quark or gluon, and find that they lead to different angular
distributions of the decay products. While quarks, at twist-2 level,
contribute to $b_1$ only, the gluons contribute both to $b_1$ and
$\Delta$.

We start with the  quark case which is
simpler, following \cite{ET82},
where the quark tensor fragmentation function was introduced and
investigated, and, in particular, a sum rule for 
its first moment was
derived, which was later obtained for the analogous 
distribution function by Close
and Kumano \cite{CK}. 

We describe the tensor polarization by a Cartesian tensor 
$S^{\mu \nu}$, which is symmetric, traceless and transverse,
and which should be considered at the same footing as the familiar vector
polarization $S^\mu$. 

\be
\label{S}
  S^{\mu \nu} = S^{\nu \mu}\;, \;\;\;\; S^{\mu}_{~\mu} = 0 \;, \;\;\;\;
S^{\mu \nu}\,P_\nu =0 \;\;.
\ee

In the rest frame the $3 \times 3$ density matrix is reduced,
apart of the trace, which describes the spin-averaged cross-section, 
to the 3 components of the polarization vector $S^i$ 
and the 5 components of the symmetric traceless tensor $S^{ij}$.

For pure polarization states the density matrix is just $\rho^{\mu\nu}=
\epsilon^{\mu}\epsilon^{*\nu}$, and its traceless part is 
\be
\label{rho}
S^{\mu \nu} = \frac{1}{2}\left(
\epsilon^{\mu}\epsilon^{*\nu} + \epsilon^{\nu}\epsilon^{*\mu} \right) -
\frac{1}{3}\left(g^{\mu \nu} - \frac{P^\mu P^\nu}{M^2}
\right)\,\epsilon^{\alpha}\epsilon^*_{\alpha} \;\;.
\ee
where $P^\mu$ and $M$ are the meson momentum and mass, respectively.

The $S^{\mu \nu}$-dependent twist-2 terms 
in the quark fragmentation function 
\cite{CS82}
are described by the following 
expression: 
\begin{eqnarray}
\label{def}
& z P^+ \int \frac{dx^-}{4\pi} \exp \left(-iP^+\frac{x^-}{z}
\right)\;\frac{1}{3}Tr_{Color}\; \frac{1}{2}Tr_{Dirac} \sum\limits_{P,X} 
 <0|
\gamma^\mu\psi (0)|P, X><P,X| \bar \psi (x)|0> &\nonumber \\ 
& = \bar f (z, \mu^2)P^\mu+
 M^2 \left(\bar f_{LT} (z, \mu^2)
S^{\mu \nu} n_\nu +\bar f_{LL} (z, \mu^2)
P^\mu S^{\rho \nu} n_\rho n_\nu \right)& 
\end{eqnarray}
where $z=\frac{P^+}{k^+}$, $\mu$ is a factorization scale, 
 $\bar f(z,\mu^2)$ is the standard spin-averaged 
term, $n_\nu$ is a light-cone vector normalized by the condition $P
\cdot n=1$,
and we introduced the two tensor fragmentation functions
$\bar f_{LT}(z, \mu^2)$ and $\bar f_{LL}(z, \mu^2)$. 

We are not considering any power corrections here, and in this limit
only the purely longitudinal component \cite{ET82} 
of the tensor $S^{\mu \nu}$ contributes.
\be
\label{SL}
S^{\mu \nu} \Rightarrow \frac{P^\mu P^\nu}{M^2}S_{zz}
\ee
where the meson is assumed to move
along the $z$ axis
in the infinite momentum frame. 
The coefficient $S_{zz}$ is the component of the tensor polarization
in the meson rest frame, when the direction of the $z$ axis is chosen
along the meson momentum in the infinite momentum frame (before the
boost to the rest frame),
Consequently only the sum of the functions $\bar f_{LT}$ and $\bar f_{LL}$
is entering, which is just $f_{Al}$ in (20) of 
\cite{ET82} and the analog of $b_1$ for the distribution case:

\be
\label{b1}
\bar b^q_1(z,\mu^2)=\bar f_{LT}(z,\mu^2) +\bar f_{LL}(z,\mu^2)
\ee

The tensor polarization dependence of a given hard 
cross-section
is obtained by the convolution of (\ref{def}) with the hard scattering 
kernel. The factor $P^\mu$ in (\ref{def}), (\ref{SL}) is playing the 
role of the 
unpolarized quark density matrix. 
Consequently, the hadron cross-section is just a convolution
of the {\it unpolarized} partonic 
cross-section with this new function 

\be
\label{dsigma}
 \frac{d\sigma}{dt} = \int dz\; \bar b^q_1(z,\mu^2)\; \frac{d\hat
\sigma}{d \hat t} (z,\mu^2)
\;\;,
\ee
where 
$ \frac{d \hat \sigma}{d \hat t}$ is the corresponding spin-averaged 
partonic cross-section.

Let us consider the analogous contribution due to gluons.
Restricting ourselves to the terms symmetric in $\mu$ and $\nu$ 
(the antisymmetric terms are related to the vector polarization which
is not discussed in this paper) and averaging
over gluons colours and spins, we get \cite{CS82}:
\begin{eqnarray}
\label{defG} 
&- \frac{z}{16 \pi P^+}\int dx^-
\exp\left(-iP^+\frac{x^-}{z}\right) \sum\limits_{P,X}
<0|G^b_{+ \mu}(0)|P,X><P,X|G^b_{+ \nu}(x)|0>& 
\nonumber \\
&=\bar G(z,\mu^2)g^{\perp}_{\mu \nu}
+\bar \Delta (z,\mu^2) S^{\perp}_{\mu \nu}+
M^2 \bar b^G_1(z,\mu^2) g^{\perp}_{\mu \nu} S_{\rho \sigma} n^\rho n^\sigma & 
\nonumber \\
\end{eqnarray}

where $g^{\perp}_{\mu \nu}=g_{\mu \nu}-g^{||}_{\mu \nu}$ with 
$g^{||}_{\mu \nu}=P_\mu n_\nu+P_\nu n_\mu-M^2n_\mu n_\nu$ and 
$G$ is the familiar spin-averaged term. The
transverse component of the polarization tensor is 

\begin{eqnarray}
\label{defS}
&&S^{\perp}_{\mu \nu}=\frac{1}{2}(g^{\perp}_{\mu \alpha} g^{\perp}_{\nu \beta}
+g^{\perp}_{\nu \alpha} g^{\perp}_{\mu \beta}-
g^{\perp}_{\mu \nu} g^{\perp}_{\alpha \beta})S^{ \alpha \beta}
\nonumber \\
&& \approx
g^{\perp}_{\mu \alpha} g^{\perp}_{\nu \beta}S^{\alpha \beta},
\end{eqnarray}
where symmetry and tracelessness were taken into account to get the
second expression, valid, when the mass corrections 
to $g^{\perp}_{\mu \nu}$
are neglected.

Note the important difference between quarks and gluons. While both are
contributing to the function $\bar b_1$ \cite{ET82,ET93}, 
it is only gluons, which provide 
the leading twist contribution to $\bar \Delta$. 
%The formal reason, 
%directly applicable for the respective distributions, is related to
%the different
%symmetry properties
%of the operators involved (we consider the simpler case of the
%distribution functions at this point): 
%\be
%\label{expl}
%<P,S|G_b^{\mu +}(0) D^{\nu_1}...  D^{\nu_n}
%G_b^{\nu +}(0) |P,S> \sim S^{\mu \nu}
%P^{\nu_1}... P^{\nu_n},
%\ee
%where $D^{\nu_i} (i=1,...,n)$ 
%is the covariant derivative in the adjoint representation.
%While the moments of $b_1$ are related to the operators which are 
%completely symmetric in $\mu, \nu, \nu_i$ 
%(the same as in the unpolarized case), $\Delta$ 
%is produced by operators of mixed symmetry, being symmetric with respect
%to the interchange of $\mu $ and $\nu$, as well as 
%to the interchange of any of the $\nu_i$, but antisymmetric
%under interchanges between the indices $\mu$ or $\nu$ and $\nu_i$. 
%For quarks, the
%operators with such a mixed symmetry do not appear at the leading
%twist level. 
Physically, the reason is that a (on-shell) gluon 
is described by a $2 \times 2$ matrix of Stokes parameters 
(in contrast to a massless quark whose polarization state is completely 
characterized by just one number, namely its helicity)
%has an additional degree 
%of linear polarization (like the photon, whose polarization state
%is described, besides the helicity, by the plane in which the 
%vectors of electric and magnetic fields are oscillating)
and thus carries additional
information on the parent hadron. 

The functions $\bar b_1$ and $\bar \Delta$ describe different 
components of the
meson tensor polarization. While $\bar b_1$ in the meson rest frame 
is related to the component $S_{zz}=-S_{xx}-S_{yy}$, $\bar \Delta$ 
is describing
another degree of freedom, namely $S_{xx}-S_{yy}$.

To come to the observable quantities one should first recall \cite{LL,ET82}
that $S^{ij}$ is actually the polarization, selected by the
detector.
By isolating specific angular correlations in the final state 
one defines $S_{ij}^{det}$. More precisely, 
the measured cross-section allows to
extract the tensor polarization 
$ S_{ij}^{scat}$ according to:
\be
\label{s_scat}
d\sigma=d\sigma_0(1 + S_{ij}^{scat} S_{ij}^{det}),
\ee
where $d\sigma_0$ is the spin-averaged cross-section.
The angular distribution for e.g. the decay into a pair of scalar
particles (pions, kaons) in the rest frame is of the form:
\be
\label{ang}
\frac{d^2\sigma}{d \theta d \phi} \sim
1 + S_{zz}^{scat}
 (3 cos^2\theta-1) + (S_{xx}^{scat}-
S_{yy}^{scat}) sin^2\theta cos 2\phi 
\ee
where $\theta$ is the polar angle with respect to the $z$ axis and  $\phi$
is the azimuthal angle with respect to the reaction plane. 
The standard choice of the $z$ axis is along the meson momentum $\vec
P$ in the target rest frame.
Both angles are measured in the vector meson rest frame.

To our knowledge, the unifying 
treatment of $b_1$ and $\Delta$ as describing  different
components of the polarization tensor is absent in the literature, although 
the appearence of $\Delta$ in the description of the difference of 
deutron cross-sections with orthogonal linear polarization 
was already presented in the first paper of \cite{JM89}.

Recall \cite{ET82}, that for $\bar b_1$, 
the short-distance subprocesses are the same as in
the unpolarized case, as they are projected onto the same structure.
At the same time, the structure $S^{\perp}_{\mu \nu}$ may be
considered as a density matrix of the linearly polarized gluons.
As a result, 
$\bar \Delta$ is picking out that 
part of the final gluon density matrix which describes 
its linear polarization.

Such partonic subprocesses were quite extensively studied in the past. 
Let us briefly discuss, which hadronic reactions are of potential importance
for us. 

The linear polarization of produced gluons may emerge in scattering
processes with
unpolarized initial particles. In particular, the gluon linear polarization
in the $q \bar q$ and $qg$ subprocesses of $pp$ scattering \cite{DR82}
is zero in the Born approximation for massless quarks and requires to
take into 
account one-loop contributions. The resulting polarization is 
sensitive to the QCD colour structure and is reasonably large (of the order
$10\%$). The
observation of a $\cos 2 \phi$ dependence in the vector mesons decays,
correlated with the predicted angular dependence of the polarization
of the scattered gluon,
would give access to $\bar \Delta$. However, 
as the resulting coefficients in the
angular distribution are not expected to be large, high statistics data 
(say, from RHIC) would be required. 

Another promising candidate is electron-positron 
annihilation\cite{OOO80}. The relevance of vector mesons 
(especially isosinglet ones, which are
supposed to be strongly coupled to gluons) was already mentioned there.
However, the related comprehensive 
analysis, explicitely specifying the relevant
fragmentation functions in a way presented here, 
is, to our knowledgs, absent in the literature.   
The last remark applies also to 
recent studies \cite{GKL98} of the linear
polarization of gluons accompanying $t \bar t$ quarks pairs.  

The gluon linear polarization in DIS is appearing due to the linear
polarization of the virtual photon. The situation is different for large and
small $Q^2$. For large $Q^2$ the gluon polarization  is calculable
perturbatively. The gluon linear polarization in the QCD Compton scattering
subprocess was studied in detail some years ago \cite{OOO84}. 
The obtained numerical results are fairly large, in particular, for HERA
kinematics. 
As it is clear from our analysis, 
the studies of the azimuthal dependences of decay products
are providing access to the tensor 
fragmentation functions. 

The angular distribution of the decay pions is given by (\ref{ang}).
The fragmentation functions $\bar b_1$ of all quarks and gluons contribute
to the polar angle dependence:
\be
\label{finzz}
S_{zz}^{scat}= \frac
{\sum\limits_{q}\int dxdz\; q(x) [\bar 
b^q_1(z)\; \frac{d\hat \sigma^q}{d \hat t}+
\bar 
b^G_1(z)\; \frac{d\hat \sigma^G}{d \hat t}]}
{\sum\limits_{q}\int dxdz\; q(x) [\bar 
f^q(z)\; \frac{d\hat \sigma^q}{d \hat t}+
\bar 
G(z)\; \frac{d\hat \sigma^G}{d \hat t}]}.
\ee
Here $q(x)$ is the standard spin-averaged 
quark distribution. We drop for brevity
here and in the following formulas the dependence on the
factorization scale which should be, as usual, of the order of the 
transverse meson momentum.  

The azimuthal angle dependence at the leading twist level is
directly related to the gluon fragmentation function 
$\bar \Delta$ (Fig.1):
\be
\label{finxx}
S_{xx}^{scat} - S_{yy}^{scat}
= \frac{
\sum\limits_{q} 
\int dxdz\; q(x) \bar 
\Delta (z)\; [\frac{d\hat \sigma^G
_{xx}}{d \hat t}-\frac{d\hat \sigma^G_{yy}}{d \hat t}]}
{\sum\limits_{q}\int dxdz\; q(x) [\bar 
f^q(z)\; \frac{d\hat \sigma^q}{d \hat t}+
\bar 
G(z)\; \frac{d\hat \sigma^G}{d \hat t}]}
\ee
where 
$\frac{d\hat \sigma_{ii}}{d \hat t}(i=x,y)$
is the partonic cross-section with the produced gluon 
being linearly polarized along the $i$ axis, so that: 
\be
\label{sig}
\frac{d\hat \sigma^G
_{xx}}{d \hat t}+\frac{d\hat \sigma^G_{yy}}{d \hat t}
=\frac{d\hat \sigma^G}{d \hat t}
\ee

One of the two integrations in the parton momentum fractions may be 
eliminated by making use of the delta-function in the parton-cross 
section. Taking also into account that the distribution
function decreases rapidly with growing $x$, which is required 
to produce the fast gluon with small $z$, we conclude that small $z$
gives only negligible contribution and restrict 
the integration to the leading hadron  
region $z \sim 1$. This corresponds to the case of the free gluon
polarization, considered in \cite{OOO84}, and which is most important 
for the experimental studies.

The experimental measurement of angular distributions allows to
determine the l.h.s. of the above equations. Developing estimates
\cite{ET82} and models \cite{b1,lat} for tensor fragmentations
functions, one may evaluate the size of the expected  experimental signal.

\begin{figure}[h]
\begin{center}
\epsfig{file=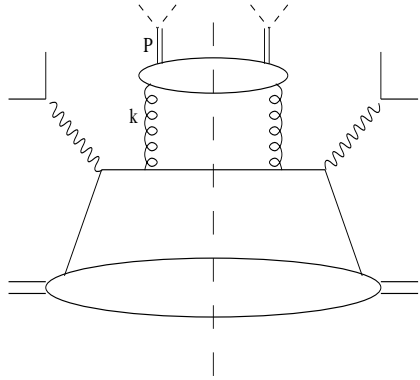, height=5cm,width=5.5cm}
\caption{Typical diagram for the high $P_T$ 
production of a $\rho$ meson in semi-inclusive DIS}
\end{center}
\end{figure}

\begin{figure}[h]
\begin{center}
\epsfig{file=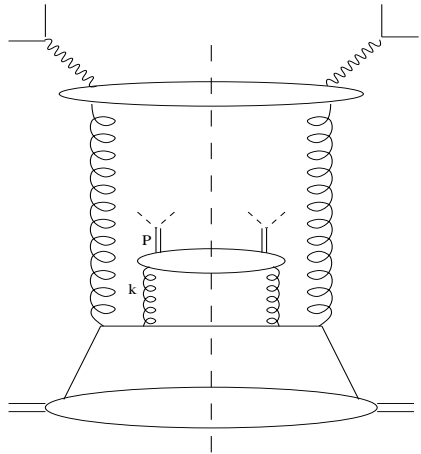, height=6cm,width=5.5cm}
\caption{Typical digram for photoproduction of a $\rho$
meson with high $P_T$}
\end{center}
\end{figure}

For small $Q^2$, there is a significant resolved 
contribution parametrized by the photon structure
function (Fig. 2).
The term, proportional to  $\bar \Delta$ is picking out, due
to helicity conservation. The double helicity flip term 
is given by the distribution of gluons in the photon 
$\Delta^\gamma$, whose convolution with the quark 
box diagram results in the structure function $F_3^\gamma$. 

\be
\label{res}
S_{xx}^{scat} - S_{yy}^{scat}
= \frac{
\sum\limits_{q} 
\int dx_g dx_q dz
\; \frac{2(1-y_\gamma)}{y_\gamma
} \Delta^\gamma (x_g) q(x_q) \bar 
\Delta (z)\; [\frac{d\hat \sigma^G
_{xx}}{dt}-\frac{d\hat \sigma^G_{yy}}{dt}]}
{\sum\limits_{q}\int dx_g dx_q dz\;
\frac{1+(1-y_\gamma)^2}{y_\gamma}
G^\gamma (x_g) q(x_q) [\bar 
f^q(z)\; \frac{d\hat \sigma^q}{dt}+
\bar 
G(z)\; \frac{d\hat \sigma^G}{dt}]},
\ee
where
$G^\gamma$ is the spin-averaged distribution of gluons in the photon, while
$y_\gamma=2\nu/s,~y_\gamma x_g$ are the fractions of the electron 
momentum, carried by the photon and gluon, respectively. 
This equation is just Eq. (\ref{finxx}) in which nominator and
denominator were convoluted with the appropriate gluon distribution. 
%The gluon momentum fraction is called $x_g$. 
Depending on the experimental situation, one might also have to
integrate over $y_\gamma$.

Several additional comments might help to better understand this formula.

Firstly, note that 
the Weizs\"acker-Williams factors differ for the linearly polarized
and unpolarized case (c.f. also \cite{JM89})
and their ratio is just the linear polarization of the emitted (almost
on-shell) photon \cite{BGMS74}:
\be
\label{xi}
\xi=\frac{2(1-y_\gamma
)}{1+(1-y_\gamma)^2}.
\ee
This formula, which may be easily recovered from, say, \cite{BGMS74},
has a simple physical interpretation. Recall that $(1 \pm 
(1-y_\gamma)^2)/y_\gamma$
are the DGLAP spin-averaged and spin-dependent kernels, respectively.
The factors $1/y_\gamma
$ and $(1-y_\gamma
)^2)/y_\gamma$ are then the
kernels for the productions of the photon of
positive and negative helicity by the electron of positive helicity, 
so that $1/\sqrt{y_\gamma
}$ and 
$(1-y_\gamma)/\sqrt{y_\gamma}$ may be understood as the respective helicity
amplitudes (up to inessential phases). The interference term, 
corresponding to the double helicity flip, is then $2(1-y_\gamma
)/y_\gamma$, and the ratio `double helicity flip over spin-averaged 
case' becomes $2(1-y_\gamma)/(1+(1-y_\gamma)^2$
in agreement with (\ref{xi}). The polarization is maximal for $y_\gamma
 \to
0$, when flip and non-flip amplitudes are equal. If the integration 
over $y_\gamma$ is performed in (\ref{res}), the contribution of this 
region is enhanced by the factor $1/y_\gamma$.

Secondly, other partonic subprocesses besides QCD Compton 
scattering (Fig. 2) contribute both in
the numerator and denominator.  
The numerator can receive an extra contribution from gluon-gluon
scattering. It is dominant at small $x$ and requires a separate
investigation following the lines of \cite{EKS99}.
In the present paper we assume $P_T$ to be large enough to neglect such 
a term. 
The denominator describing the spin averaged cross-section
receives an additional contribution from all possible combinations of 
quarks (antiquarks distributions are assumed to be small) and gluons. 
At the same time, the quark-quark subprocess dominates at large $x$,
corresponding to very large $P_T$ and small cross-sections, while the 
gluon-gluon scattering is dominant at small $x$. So,  
one should add only the convolution of 
the quark distribution 
in the photon with the gluon distribution in the nucleon 
neglected just 
for brevity. 

Thirdly, the analyzing power of the short distance 
production of linearly polarized gluons by linearly 
polarized photons
by the QCD Compton  
subprocess was found
\cite{PP80} to be fairly large.
The analysing power for the QCD Compton 
subprocess of elastic scattering of linearly polarized gluons
is actually the same,
as it is controlled only by chirality and angular momentum
conservation.  
%The actual correlation for the subprocesses is
%obtained by a kinematical transformation, common for photon and
%gluon. 

Let us conclude: we have analyzed how to extract from the angular 
distribution of $\rho$ and other vector mesons produced in DIS 
of unpolarized electrons on unpolarized nucleons and nuclei the novel
fragmentation functions $\bar b_1^q(z), \bar b_1^G(z)$ 
(from Eq. (\ref{finzz})), $\bar \Delta(z)$ 
(from Eq. (\ref{finxx})) and the photon
distribution
$\Delta^\gamma(x)$, directly related to the 
structure function
$F_3^\gamma(x)$, from photoproduction
(Eq. (\ref{res})).
\vskip.3in
%\newpage
{ \Large \bf Acknowledgements:}\\
\vskip.2in
We acknowledge the useful 
discussions with J.~Crittenden, A.V.~Efremov, R.L.~Jaffe, B.~Pire, 
D.H.~Schiller and M.I.~Strikman. \\
A.S. and L.S. acknowledge the support from DFG and BMBF.
O.V.T. was supported by Erlangen-Regensburg Graduiertenkollegs and by DFG in
the framework of Heisenberg-Landau Program of JINR-Germany Collaboration.

%\newpage

\bb{99}
\bi{HJM89} P. Hoodbhoy, R.L. Jaffe and A. Manohar, Nucl. Phys.
B312(1989)571; 

\bi{FS}
L.L. Frankfurt and M.I. Strikman, Nucl.Phys. A405(1983)55; 

\bi{b1} N.N. Nikolaev and W. Sch\"afer, Phys. Lett. B398(1997)245;\\
K. Bora and R.L. Jaffe Phys. Rev. D57(1998)6906;

\bi{JM89} R.L. Jaffe and A. Manohar, Phys. Lett. B223(1989)218;\\
X. Artru and M. Mekhfi, Zeit. f. Physik C45(1990)669; \\ 
E. Sather and C. Schmidt, Phys. Rev. D42(1990)1424;

\bi{lat} C. Best {\it et al},  Phys. Rev. D56(1997)2743;

\bi{BKLF85}
A.P. Bukhvostov, G.V. Frolov, L.N. Lipatov, E.A. Kuraev, 
Nucl.Phys. B258(1985)601;

\bi{EKS99} B. Ermolaev, R. Kirschner and L. Szymanowski, Eur. Phys. J.
C7(1999)65;

\bi{exp} P. Neuman, "Difractive Phenomena at HERA", hep-ex/9901026; 
   \\
P. Abreu {\it et al} (DELPHI Coll.),  Phys. Lett. 406B(1997)271.

\bi{ET82} A.V. Efremov and O.V. Teryaev, Sov.J.Nucl.Phys. 36(1982)557; 

\bi{CK} F.E. Close and S. Kumano, Phys. Rev. D42(1990)2377;

\bi{CS82} J.C. Collins and D.E. Soper, Nucl. Phys. B194(1982)445; 

\bi{ET93} A.V. Efremov and O.V. Teryaev,
On the oscillations of the tensor spin structure function,
Proc. of Int. Symp. "DUBNA DEUTERON-93", JINR E2-94-95, pp.206-210.

\bi{LL} W.B. Berestetzki, E.M. Lifshitz and L.P. Pitajewski, 
Quantum Electrodynamics, Nauka, Moscow, 1980;

\bi{DR82} A. Devoto and W.W. Repko, Phys. Lett. 25B(1982)904;

\bi{OOO80} H.A. Olsen, P. Osland and I. {\O}verb{\o}, Phys. Lett.
89B(1980)221, Nucl. Phys. B171(1980)209;  

\bi{GKL98} S. Groote, J.G. K{\"o}rner and J.A. Leyva, Eur. Phys. J.
C7(1999)49;

\bi{PP80} B. Petersson and B. Pire, Phys. Lett. 95B(1980)119;

\bi{OOO84} {\O}.E. Olsen and H.A. Olsen, Physica Scripta, 29(1984)12;

\bi{BGMS74} V.M. Budnev, I.F. Ginzburg, G.V. Meledin and V.G. Serbo,
 Phys. Rept. C15(1975)181, Chapter 6.6.

\eb

\end{document}